\documentclass[12pt]{iopart}
\usepackage[utf8]{inputenc}
\usepackage[english]{babel}
\usepackage{lmodern}
\usepackage{graphicx}
\usepackage{epstopdf}
\usepackage{upgreek}
\usepackage{color}

\begin{document}

\title[Jaynes-Cummings-Rabi model and charged Dirac particle]{Quasienergy collapse in the\\ driven Jaynes--Cummings--Rabi model:\\ correspondence with a charged Dirac particle\\ in an electromagnetic field}
\author{R. Guti\'{e}rrez-J\'{a}uregui and H. J. Carmichael$^{\dagger}$}
\address{The Dodd-Walls Centre for Photonic and Quantum Technologies,
Department of Physics, University of Auckland,
Private Bag 92019, Auckland, New Zealand}
\ead{$^{\dagger}$h.carmichael@auckland.ac.nz}


\begin{abstract}
A system evolving under the driven Jaynes--Cummings model will undergo a phase transition at a critical driving field amplitude. This transition is foreshadowed by a collapse of the quasienergy level spectra of the system and remains present as the model is extended to include a counter-rotating interaction. We study this critical response and obtain the eigenvalues and eigenstates of the extended model by presenting a correspondence between the Jaynes--Cummings model and a charged Dirac particle subject to an external electromagnetic field. Under this correspondence, the field and two-level system that compose the Jaynes--Cummings model map to the external and internal degrees of freedom describing the Dirac particle, respectively. The phases of the system below (above) the critical drive are then characterized by discrete (continuous) solutions, with the manipulations required to obtain these solutions appearing naturally as Lorentz transformations.

\end{abstract}
\noindent{\it Keywords\/ }: driven Jaynes--Cummings-Rabi model, Dirac equation, quantum phase transition

\maketitle

\section{Introduction}

The interaction between a two-level system and a harmonic oscillator is ubiquitous in modern physics. This universality allows us to extrapolate ideas and techniques across different physical backgrounds which---appealing to the developed intuition---may appear more natural on one field over another. In a quantum optical context, the harmonic oscillator and the two-level system compose the quintessential model for the quantum theory of radiation and its interaction with matter, with the former modelling a single electromagnetic field mode and the latter describing two discrete energy levels of a material system. The dynamics of the coupled system are described up to a very good approximation by the Jaynes--Cummings (JC) model if the coupling is dipolar in character and the rotating wave approximation is valid~\cite{Jaynes_1963}. While deceptively simple, these dynamics allow for deep physical insight and conceptual understanding of quantum concepts, as exemplified in the experimental demonstration of the discrete nature of the electromagnetic field~\cite{Meekhof_1996,Brune_1996,Hofheinz_2008}.

The operational simplicity of the model seems to translate to the driven Jaynes--Cummings model. The driven JC model describes a two-level system interacting with a field mode that is being driven out of the ground state by an external coherent field. When the resonance frequency of the two-level transition, field mode frequency, and drive frequency share the same value, it is possible to move to an interaction picture where the dynamics are simplified. In this interaction picture, the evolution of the system is ruled by the time-independent Hamiltonian
\begin{equation}\label{back_1}
\mathcal{H}_{0} = i\hbar \lambda \left( \hat{a} \hat{\sigma}_{+} - \hat{a}^{\dagger} \hat{\sigma}_{-} \right) + \hbar \epsilon \left( \hat{a} + \hat{a}^{\dagger} \right) \, ,
\end{equation}
with $\hat{a}^{\dagger}$ and $\hat{a}$ the creation and annihilation operators of the field mode, $\hat{\sigma}_{+}$ and $\hat{\sigma}_{-}$ the raising and lowering operators of the two-level system, and parameters $\lambda$, $\epsilon$ to determine the dipole coupling strength and external field amplitude, respectively. This model describes an idealized scenario where a phase transition of light is found~\cite{Carmichael_2015,Alsing_1992_2,Alsing_1992_1}, caused by the competition between coupling strength and driving field amplitude, with the former advocating for a definite excitation number while the latter advocates for a definite phase. The phase transition is characterized by a change in the level structure of the Hamiltonian, equation~(\ref{back_1}), at the critical drive amplitude
\begin{equation}\label{back_2}
\epsilon_{0} = \lambda / 2 \, .
\end{equation}
For driving amplitudes below $\epsilon_{0}$ the Hamiltonian displays a discrete level structure, with level spacings that collapse to zero when this critical value is reached; the level structure becomes continuous above it~\cite{Alsing_1992_2}.

The critical behaviour persists if the interaction between the two-level system and field mode is extended to include counter-rotating coupling~\cite{Gutierrez_2017b}. In this extended model the dynamics in the interaction picture are ruled by the generalized Jaynes--Cummings--Rabi Hamiltonian
\begin{eqnarray}\label{back_3}
\mathcal{H}_{\eta} = i\hbar \lambda^{\prime} \left[ \left( \hat{a} + \eta \hat{a}^{\dagger} \right) \hat{\sigma}_{+} - \left( \hat{a}^{\dagger} + \eta \hat{a} \right) \hat{\sigma}_{-} \right]
+ \hbar \epsilon^{\prime} \left( \hat{a} + \hat{a}^{\dagger} \right) \, ,
\end{eqnarray}
with $\eta$ a continuous parameter ranging from $\left[ 0 ,1 \right]$. The expected dynamics under this Hamiltonian can be realized using a cavity QED architecture~\cite{Dimer_2007}.
In this realization the rotating and counter-rotating interactions are controlled independently by promoting externally driven Raman transitions between a pair of ground states via a manifold of far-detuned excited states; the time-independent Hamiltonian, equation~(\ref{back_3}), is realized in an appropriate interaction picture (see Sections IIIC and D of Reference~\cite{Gutierrez_2017b}). Considering this open system perspective clarifies the origin of the photons injected into the field mode through the counter-rotating coupling.
In fact, equations~(\ref{back_1}) and~(\ref{back_3}) are connected through a squeezing transformation,
\begin{equation}\label{back_6}
\mathcal{S}(z) \mathcal{H}_{0} \mathcal{S}^{\dagger}(z) = \mathcal{H}_{\eta} \, ,
\end{equation}
with squeeze parameter given by
\begin{eqnarray}\label{back_7_t}
\cosh z = \lambda^{\prime}/ \lambda \, , \label{back_7_1} \\ \sinh z =  \eta \lambda^{\prime} / \lambda   \, , \label{back_7_2}
\end{eqnarray}
which introduces the scaled driving field amplitude
\begin{equation}\label{back_7_3}
\epsilon^{\prime} = \epsilon ( \lambda^{\prime} / \lambda ) \left( 1 + \eta \right) \, .
\end{equation}
The addition of the counter-rotating interaction not only affects the mode photon number, but also displaces the critical drive towards the value
\begin{equation}\label{back_4}
\epsilon^{\prime}_{\eta} =  \lambda^{\prime} (1 + \eta) / 2 \, .
\end{equation}
These transformations establish a simple connection between the on-resonance JC and Jaynes--Cummings--Rabi models. We emphasize, in particular, that they allow us to target the treatment in the following sections to the JC interaction without loss of generality.

Previously the quasienergies and eigenstates of the system below the critical amplitude have been obtained through a Bogoliubov transformation~\cite{Alsing_1992_2,Gutierrez_2017b}; this method ceases to be applicable, however, for the continuous spectrum above the critical amplitude. We overcome this problem in the following, where we develop a new method for obtaining the quasienergies and eigenstates, both below and above critical drive, for the driven Jaynes--Cummings model. The solutions of the generalized model are then readily available through the squeezing transformation [equations~(\ref{back_6})-(\ref{back_7_3})]. The method is based on the correspondence between this model and the dynamics of a Dirac particle subject to an appropriate electromagnetic field configuration. Once the correspondence is established, the Lorentz covariance of the Dirac equation will be exploited to reach the desired solutions.

Section~\ref{secc:Dirac} is devoted to the correspondence between the relativistic and quantum optical models. We first introduce the Dirac equation for a charged particle subject to an external electromagnetic field and then present a particular field configuration where the evolution of the Dirac particle mimics the system under the driven Jaynes--Cummings Hamiltonian. In Section~\ref{secc:Lorentz} we obtain the eigenvalues and eigenstates of the Dirac equation; by considering the relativistic invariants of the electromagnetic field, they are shown to display two distinct behaviours which are connected to particularly simple physical scenarios through Lorentz transformations. In Section~\ref{secc:Jaynes_cummings} we return to the quantum optical model and review the solutions in this context, where the relation between Lorentz and squeezing transformations is discussed. Section~\ref{secc:conclusion} is left for conclusions.

\section{Correspondence to a relativistic system}\label{secc:Dirac}

The dynamics of a spin-1/2 particle of mass $m$ and charge $e$ in the presence of an external electromagnetic field are accurately described by the Dirac equation
\begin{equation}\label{corr_1}
\left[ \gamma^{\mu}\left( \hat{p}_{\mu} - \frac{e}{c}A_{\mu} \right) - mc \right] \psi_{D} = 0 \, ,
\end{equation}
where $\hat{p}_{\mu}=  i \hbar \partial_{\mu}$ is the four-momentum operator,  $A_{\mu}$ the electromagnetic potential, and the covariant $\gamma$ matrices satisfy the anticommutation relation
\begin{equation}\label{corr_2}
\left\lbrace \gamma^{\mu}, \gamma^{\nu} \right\rbrace_{+} = 2 g^{\mu \nu} \, ,
\end{equation}
with $g^{\mu \nu}$ the metric tensor. We adopt the metric tensor with signature $\left( 1 , -1, -1, -1 \right)$ and consider the symmetric representation of the Dirac matrices:
\begin{eqnarray}\label{corr_3}
 & \gamma^{0} = \left( \begin{array}{cc}
0 & 	\mathbf{1}_{2 \times 2}\\
\mathbf{1}_{2 \times 2} & 0 \\
\end{array} \right) 	 \, , \hspace{.025\linewidth}
\gamma^{i} = \left( \begin{array}{cc}
0 & 	\sigma^{i}\\
-\sigma^{i} & 0 \\
\end{array} \right)  \, ,
\end{eqnarray}
written in terms of the Pauli matrices $\sigma^{i}$.

Consider the scenario of a massless Dirac particle $(m=0)$ subject to constant and homogeneous electric, $\mathbf{E} = E {\mathbf{e}}_{1}$, and magnetic, $\mathbf{B} = - B {\mathbf{e}}_{3}$, fields. This field configuration is commonly used for measuring magnetoresistance and transport properties of electron gases; relevant examples are found in~\cite{MacDonald_1983, Lukose_2007} where the electron gas is effectively described by equation~(\ref{corr_1}). The electromagnetic potential generating this field configuration is determined up to a gauge transformation, with the gauge commonly selected to match the boundary conditions of the problem at hand~\cite{Champel_2007}. We select
\begin{equation}\label{corr_4}
A_{\mu} = \left(E x_{1}; 0 , - B x_{1}, 0 \right) \, ,
\end{equation}
for which equation~(\ref{corr_1}) becomes invariant under translations in the ${\mathbf{e}}_{2}$ and ${\mathbf{e}}_{3}$ directions giving rise to the conserved quantities $\hat{p}_{2}$ and $\hat{p}_{3}$. Solutions to equation~(\ref{corr_1}) can be written in terms of $\phi^{+}$ and $\phi^{-}$ spinors through
\begin{equation}\label{corr_5}
\psi_{D} = \left( \begin{array}{c}
\phi^{+}_{D}\\
\phi^{-}_{D} \\
\end{array} \right) 	 \, ,
\end{equation}
which, for $m=0$, satisfy the uncoupled equations
\begin{eqnarray}
\left\lbrace i \hbar \partial_{t} - e E x_{1} + c \sigma_{1}\hat{p}_{1} + \sigma_{2}eBx_{1} + c \sigma_{j} \hat{p}_{j} \right\rbrace \phi^{+}_{D} =0 \, , \label{corr_6a} \\
\left\lbrace i \hbar \partial_{t} - e E x_{1} - c \sigma_{1}\hat{p}_{1} - \sigma_{2}eBx_{1} -c \sigma_{j} \hat{p}_{j} \right\rbrace \phi^{-}_{D} =0 \, , \label{corr_6b}
\end{eqnarray}
with $j=2,3$ and the Einstein summation convention: $\sigma_{j} \hat{p}_	{j} = \sigma_{2} \hat{p}_{2} + \sigma_{3} \hat{p}_{3}$ is understood. The potential generated through the interaction with the magnetic field allows the position and momentum operators to be written in terms of bosonic operators,
\begin{equation}\label{corr_7}
\hat{a} = \frac{x_{1} + i(l^{2}_{\mathcal{B}}/\hbar)\hat{p}_{1}}{\sqrt{2} l_{\mathcal{B}}},\hspace{.025\linewidth} \hat{a}^{\dagger} = \frac{x_{1} - i(l^{2}_{\mathcal{B}}/\hbar)\hat{p}_{1}}{\sqrt{2} l_{\mathcal{B}}} \, ,
\end{equation}
with the magnetic length
\begin{equation}\label{corr_8}
l_{\mathcal{B}} = \sqrt{\frac{\hbar c}{e B}} \,
\end{equation}
providing a natural unit to measure the depth of the potential. Equations~(\ref{corr_6a}) and~(\ref{corr_6b}) take the form
\begin{equation}\label{corr_9}
\lbrace \mp i  \hbar  \partial_{t}  - c \sigma_{j} \hat{p}_{j} \rbrace \phi^{\pm}_{D} = \left\lbrace i\sqrt{2}eB l_{\mathcal{B}} \left( \hat{a} \sigma_{+} - \hat{a}^{\dagger} \sigma_{-} \right) \mp \frac{eE l_{\mathcal{B}}}{ \sqrt{2}} \left( \hat{a} + \hat{a}^{\dagger} \right)  \right\rbrace \phi^{\pm}_{D} \,
\end{equation}
under these definitions. Since the eigenvalues of $\hat{p}_{2}$ and $\hat{p}_{3}$ remain constant through the evolution, we can consider the trivial scenario where they are zero, in which case the correspondence between equation~(\ref{corr_9}) and the evolution under equation~(\ref{back_1}) becomes transparent. The motion and spin state of the Dirac particle will map to the field mode and two-level system, respectively, in the quantum optical analogue. A similar correspondence between the undriven Jaynes--Cummings model and the Dirac oscillator was found in Ref.~\cite{Rozmej_1999}.

\section{Eigenstates and eigenvalues: Quasienergy collapse}\label{secc:Lorentz}

A method that has proven useful to determine the solutions of the Dirac equation in the presence of electromagnetic fields, equation~(\ref{corr_1}), relies on considering the Dirac-Pauli equation as an auxiliary equation~\cite{Landau_1971}. This allows us to extract a basis set for the spinors while giving a differential equation for the eigenfunctions to satisfy~\cite{Gutierrez_2017}. For a massless particle ($m=0$), the Dirac-Pauli equation takes the simplified form
\begin{equation}\label{eige_1}
\left[ \left(\hat{p}^{\mu} - \frac{e}{c} A^{\mu} \right) \left(\hat{p}_{\mu} - \frac{e}{c} A_{\mu} \right) -\frac{i e \hbar}{4c} \left[ \gamma^{\mu}, \gamma^{\nu} \right]_{-} F_{\mu \nu}  \right] \psi_{P} = 0 \, ,
\end{equation}
with
\begin{equation}\label{eige_2}
F_{\mu \nu} = \partial_{\mu} A_{\nu} - \partial_{\nu} A_{\mu} \,
\end{equation}
the electromagnetic tensor. Solutions $\psi_{D}$ to the Dirac equation are then built from solutions $\psi_{P}$ of this auxiliary second-order equation by applying the Dirac operator:
\begin{equation}\label{eige_0}
\psi_{D} = \gamma^{\mu}\left( \hat{p}_{\mu} - \frac{e}{c}A_{\mu} \right) \psi_{P} \, .
\end{equation}
For the physical scenario presented above, the relativistic invariants
\begin{eqnarray}\label{eige_3}
F_{\mu \nu} F^{\mu \nu} = \mathbf{B} \cdot \mathbf{B} - \mathbf{E} \cdot \mathbf{E} \, , \\
\epsilon_{\mu \nu \rho \eta} F^{\mu \nu} F^{\rho \eta} = 2 \mathbf{E} \cdot \mathbf{B} \, ,
\end{eqnarray}
will allow us to reduce the solutions of the system to only two cases, which correspond to the possible phases of light encountered in the quantum optical system. Given that when $F_{\mu \nu}F^{\mu \nu}$ is positive (negative) the existence of a reference frame where the electric (magnetic) field vanishes is guaranteed, the exact solution of the Dirac equation can be obtained in this privileged frame. Afterwards, by applying a Lorentz transformation, the solutions in the laboratory frame are found.

The boost required to reach the privileged frame in both scenarios must be taken in the $\mathbf{e}_{2}$ direction. The eigenvalues of $\hat{p}_{3}$ then keep the same constant value in both reference frames. Since we are interested in the solutions to the driven Jaynes--Cummings model on resonance, this operator can be considered zero throughout the following derivation without affecting the final result. Notice, however, that non-zero eigenvalues of $\hat{p}_{3}$ correspond to a non-zero detuning between the two-level system and field mode in the optical analogy, through equation~(\ref{corr_9}). We then include this operator for completeness and adopt the notation $k_{\perp} x_{\perp} = k_{2} x_{2} + k_{3} x_{3}$ where required.

\subsection{Below critical drive: Discrete spectrum}

In the case where $\vert B \vert \geq \vert E \vert$ a boost in the $\mathbf{e}_{2}$ direction with parameter
\begin{equation}\label{eige_4}
\beta_{\mathcal{B}} = \frac{E}{B} \,
\end{equation}
takes us to a reference frame where the electric field vanishes. Since the Dirac-Pauli equation is second-order in time, both Pauli spinors satisfy
\begin{equation}\label{eige_5}
\left[ \hat{p}_{0}^{\prime 2} -  \hat{p}_{1}^{\prime 2} - \left( \hat{p}^{\prime}_{2}  + \frac{eB'}{c} x^{\prime}_{1} \right)^{2} - \hat{p}^{\prime 2}_{3} - \frac{e\hbar B'}{c} \sigma_{3}  \right] \phi_{P}^{\prime} = 0 \,
\end{equation}
in this privileged frame. The basis set for the eigenspinors is then given by the eigenvectors of $\sigma_{3}$; the orientation of each is responsible for the addition of a constant in the differential equation, leaving the overall structure unaffected. Thus the differential equation is separable in all four components, with solutions given by the product of free particle states for $x^{\prime}_{2}$ and $x^{\prime}_{3}$ and eigenstates of the displaced harmonic oscillator, $H_{n}$, for $x^{\prime}_{1}$. The explicit expressions for the eigenspinors are
\numparts
\begin{eqnarray}
\phi_{P \uparrow}^{\prime} &= e^{i  k_{\perp}' x_{\perp}' }  \left( \begin{array}{c}
H_{\vert n \vert} \left( x_{1}'/l_{\mathcal{B}}'+ l_{\mathcal{B}}' k_{2}' \right)  \\
0 \\
\end{array} \right) \, , \label{eige_6a}\\
\phi_{P \downarrow}^{\prime} &= e^{i  k_{\perp}' x_{\perp}' } \left( \begin{array}{c}
0 \\
H_{\vert n \vert +1} \left( x_{1}'/l_{\mathcal{B}}'+ l_{\mathcal{B}}' k_{2}' \right) \\
\end{array} \right) \, , \label{eige_6b}
\end{eqnarray}
\endnumparts
with the corresponding Dirac particle energies given by
\begin{equation}\label{eige_7}
\upvarepsilon_{n,\pm}^{ \prime } = \pm\left( \frac{\hbar c} { l_{\mathcal{B}}'} \right) \sqrt{2 \vert n \vert + 2  + (l_{\mathcal{B}}' k^{\prime}_{3}  )^{2}} \, ,
\end{equation}
the eigenvalues of $\hat{p}_{0}^{\prime}$. The quantized motion described by these solutions is caused by the harmonic potential created through the interaction between particle spin and magnetic field. The Dirac spinors are obtained using equation~(\ref{eige_0}) and will be given by a linear combination of Pauli spinors:
\numparts
\begin{eqnarray}
\phi_{D}^{+ \prime} & = e^{i  k_{\perp}' x_{\perp}'}  \left( \begin{array}{c}
i H_{\vert n \vert} \left( x_{1}'/l_{\mathcal{B}}'+ l_{\mathcal{B}}' k_{2}' \right)  \\
c^{\prime}_{\pm} H_{\vert n \vert +1 } \left( x_{1}'/l_{\mathcal{B}}'+ l_{\mathcal{B}}' k_{2}' \right) \\
\end{array} \right) \, , \label{eige_8a}\\
\phi_{D}^{- \prime} & = e^{i  k_{\perp}' x_{\perp}' }  \left( \begin{array}{c}
c^{\prime}_{\pm}  H_{\vert n \vert} \left( x_{1}'/l_{\mathcal{B}}'+ l_{\mathcal{B}}' k_{2}' \right)  \\
i H_{\vert n \vert +1 } \left( x_{1}'/l_{\mathcal{B}}'+ l_{\mathcal{B}}' k_{2}' \right) \\
\end{array} \right) \, , \label{eige_8b}
\end{eqnarray}
\endnumparts
with the relative weights depending on the energy and momentum through
\begin{equation}\label{eige_9}
c^{\prime}_{\pm} = \pm \sqrt{\frac{\upvarepsilon_{n,\pm}^{\prime} + \hbar c k_{3}^{\prime}}{\upvarepsilon_{n,\pm}^{\prime} - \hbar c k_{3}^{\prime}}} \, .
\end{equation}
Finally, the solutions are transformed back to the laboratory frame, where the energies are found to be
\begin{equation}\label{eige_10}
\upvarepsilon_{n, \pm}=  \hbar c \left[ \beta_{\mathcal{B}} k_{2}  \pm \sqrt{\frac{\left(1 - \beta_{\mathcal{B}}^{2} \right)^{3/2}}{l^{2}_{\mathcal{B}}} \left(2 \vert n \vert + 2 \right)  +  k_{3}^{2}} \right] ,
\end{equation}
while the eigenstates are
\numparts\label{eige_11t}
\begin{eqnarray}
\phi_{D}^{+}  &= e^{i k_{\perp} x_{\perp}} \left( \begin{array}{c} i \Xi_{\mathcal{B}}^{+} H_{\vert n \vert} (z) - ic_{\pm} \Xi_{\mathcal{B}}^{-} H_{\vert n \vert +1} (z)  \\
c_{\pm} \Xi_{\mathcal{B}}^{+} H_{\vert n \vert +1 } (z) - \Xi_{\mathcal{B}}^{+} H_{\vert n \vert} (z)      \\
\end{array} \right) , \label{eige_11a}\\
\phi_{D}^{-}  &= e^{i k_{\perp} x_{\perp}} \left( \begin{array}{c}  \Xi_{\mathcal{B}}^{-} H_{\vert n \vert +1} (z) -
\Xi_{\mathcal{B}}^{+} c_{\pm} H_{\vert n \vert} (z)  \\
i \Xi_{\mathcal{B}}^{-} c_{\pm} H_{\vert n \vert} (z) - i \Xi_{\mathcal{B}}^{+} H_{\vert n \vert +1 } (z)  \\
\end{array} \right) , \label{eige_11b}
\end{eqnarray}
\endnumparts
with the displaced variable
\begin{eqnarray}\label{eige_12}
z = (1 & -\beta_{\mathcal{B}}^{2})^{1/4} \left(x_{1}/l_{\mathcal{B}} + l_{\mathcal{B}} k_{2} \right) \nonumber \\ &\pm \beta_{\mathcal{B}} \sqrt{2 \vert n \vert + 2 +  \left(1-\beta^{2}_{\mathcal{B}} \right)^{-3/2} (l_{\mathcal{B}} k_{3})^{2}} ,
\end{eqnarray}
and relative weights given by transforming equation~(\ref{eige_9}) using equation~(\ref{eige_10}) and defining the Lorentz parameter
\begin{equation}\label{eige_13}
\Xi^{\pm}_{\mathcal{B}} = \sqrt{1 \pm \left[1 - \beta_{\mathcal{B}}^{2} \right]^{1/2}}  \, .
\end{equation}
For a given $k_{2}$ and $k_{3}$, the energies of the particle display a $\sqrt{n}$-dependence that is maintained as we transform between reference frames; the separation between adjacent energies is, however, scaled by a factor accounting for dilation or contraction of the magnetic length. 

\subsection{Above critical drive: Continuous spectrum}

The same procedure can be used for $\vert E \vert \geq \vert B \vert$, where a boost in the $\mathbf{e}_{2}$ direction with parameter
\begin{equation}\label{eige_14}
\beta_{\mathcal{E}} = \frac{B}{E} \, ,
\end{equation}
takes us to a reference frame where the magnetic field vanishes. The corresponding Dirac-Pauli equation then divides into two blocks that lead to the uncoupled equations
\begin{equation}\label{eige_15}
\left[\left(\hat{p}_{0}^{\prime} - \frac{e E^{\prime}}{c} x^{\prime}_{1} \right)^{2} - \hat{p}^{\prime 2}_{1} - \hat{p}^{\prime 2}_{2} - \hat{p}^{\prime 3}_{2} \mp  i\frac{e \hbar  E^{\prime}}{c} \sigma_{1} \right] \phi^{\pm \prime }_{P} = 0 \, ,
\end{equation}
and the basis set for the eigenspinors is given by the eigenvectors of $\sigma_{1}$. While the general structure of the differential equations is similar to the previous case, there are key differences. Notice first that the energy has become a continuous parameter that displaces the origin of $x_{1}^{\prime}$. The eigenfunctions are still given by a product of free particle states for the $x_{2}^{\prime}$ and $x_{3}^{\prime}$ components, but the $x_{1}^{\prime}$ component satisfies Weber's differential equation and, as such, the corresponding eigenfunctions are given by parabolic cylinder functions
$D_{a}$~\cite{Abramowitz_1972}. The eigenspinors are
\numparts
\begin{eqnarray}
\phi_{P1}^{\pm \prime} &= e^{i k_{\perp}^{\prime} x_{\perp}^{\prime}} D_{-a-1}\left( \xi^{\prime} \right)\left( \begin{array}{c}
1 \\
\pm 1 \\
\end{array} \right) \, , \label{eige_16a} \\
\phi_{P2}^{\pm \prime} &= e^{i k_{\perp}^{\prime} x_{\perp}^{\prime}} D_{-a}\left( \xi^{\prime} \right)\left( \begin{array}{c}
1 \\
\mp 1 \\
\end{array} \right) \, , \label{eige_16b}
\end{eqnarray}
\endnumparts
where
\begin{eqnarray}
l_{\mathcal{E}}^{\prime} & = \sqrt{\frac{\hbar c}{  e E^{\prime}}} \, , \\
\xi' &=  \sqrt{2} e^{i \pi /4} \left( \upvarepsilon' l_{\mathcal{E}}^{\prime}/\hbar c - x_{1}^{\prime}/l_{\mathcal{E}}^{\prime} \right)\, , \label{eige_17}\\
a' &= -i l_{\mathcal{E}}^{\prime 2} \left( k_{2}^{\prime 2} + k_{3}^{\prime 2} \right) /2 \, . \label{eige_18}
\end{eqnarray}
The Weber differential equation is related to the harmonic oscillator equation by flipping the sign of the potential, a change that leads to imaginary values in the definition of the Pauli spinors. The idea of an effective potential that is unbounded from below can be traced back to the classical dynamics, where a charged particle subject to a constant electric field would accelerate without boundary.
The Dirac spinors corresponding to equations~(\ref{eige_16a}) and~(\ref{eige_16b}) are proportional to
\numparts
\begin{eqnarray}
\phi_{D}^{+ \prime} &= e^{i k^{\prime}_{\perp} x^{\prime}_{\perp}} \left( \begin{array}{c}
  b_{+}^{\prime}  D_{-a^{\prime}-1}(\xi^{\prime}) - D_{-a^{\prime}}(\xi^{\prime}) \\
  b_{+}^{\prime}  D_{-a^{\prime}-1}(\xi^{\prime}) + D_{-a^{\prime}}(\xi^{\prime}) \\
\end{array} \right) \, , \label{eige_19a} \\
\phi_{D}^{- \prime} &= e^{i k^{\prime}_{\perp} x^{\prime}_{\perp}} \left( \begin{array}{c}
 D_{-a^{\prime}}(\xi^{\prime}) + b_{-}^{\prime}  D_{-a^{\prime}-1}(\xi^{\prime}) \\
 D_{-a^{\prime}}(\xi^{\prime}) - b_{-}^{\prime}  D_{-a^{\prime}-1}(\xi^{\prime}) \\
\end{array} \right) \, , \label{eige_19b}
\end{eqnarray}
\endnumparts
with weights
\begin{equation}
b_{\pm}^{'} = l^{\prime}_{\mathcal{E}} e^{i \pi /4} (k_{3}^{\prime} \pm i k^{\prime}_{2})/\sqrt{2} \, . \label{eige_20}
\end{equation}

In the laboratory reference frame, the continuous spectrum is maintained and, up to a global phase factor, the eigenstates are given by
\numparts
\begin{eqnarray}\label{eige_21t}
\phi_{D}^{+} &=e^{i k_{\perp} x_{\perp}}  \left[ \Xi_{\mathcal{E}}^{+} + \Xi_{\mathcal{E}}^{-} \sigma_{2} \right] \left( \begin{array}{c}
  b_{+}  D_{-a-1}(\xi) - D_{-a}(\xi) \\
  b_{+}  D_{-a-1}(\xi) + D_{-a}(\xi) \\
\end{array} \right)  \, , \label{eige_21a}\\
\phi_{D}^{-} &=e^{i k_{\perp} x_{\perp}}  \left[ \Xi_{\mathcal{E}}^{+} - \Xi_{\mathcal{E}}^{-} \sigma_{2} \right] \left( \begin{array}{c}
 D_{-a}(\xi) -  b_{-}D_{-a-1}(\xi) \\
 D_{-a}(\xi) +  b_{-}D_{-a-1}(\xi) \\
\end{array} \right) \, , \label{eige_21b}
\end{eqnarray}
\endnumparts
with the transformed variables
\begin{eqnarray}
& \xi = \sqrt{2} e^{i \pi /4} \left(1- \beta_{\mathcal{E}}^{2} \right)^{1/4} \left[  \frac{\left( \beta_{\mathcal{E}} \hbar c k_{2} -  \upvarepsilon \right) l_{\mathcal{E}}}{  \hbar c \left(1 - \beta_{\mathcal{E}}^{2} \right) } - \frac{x_{1}}{l_{\mathcal{E}}} \right] \, , \label{eige_22} \\
& a = -\frac{i l_{\mathcal{E}}^{2}}{2} \left[ \frac{\left( k_{2} - \beta_{\mathcal{E}} \upvarepsilon / \hbar c \right)^{2}}{(1- \beta_{\mathcal{E}}^{2})^{3/2}} + \frac{k^{2}_{3}}{(1 - \beta_{\mathcal{E}}^{2})^{1/2}} \right] \, , \\
& \Xi^{\pm}_{\mathcal{E}} = \sqrt{1 \pm \left[1 - \beta_{\mathcal{E}}^{2} \right]^{1/2}}  \, . \label{eige_23}
\end{eqnarray}

\section{Back to light}\label{secc:Jaynes_cummings}

\begin{figure}[h]
\begin{center}
\includegraphics[width=.5\linewidth]{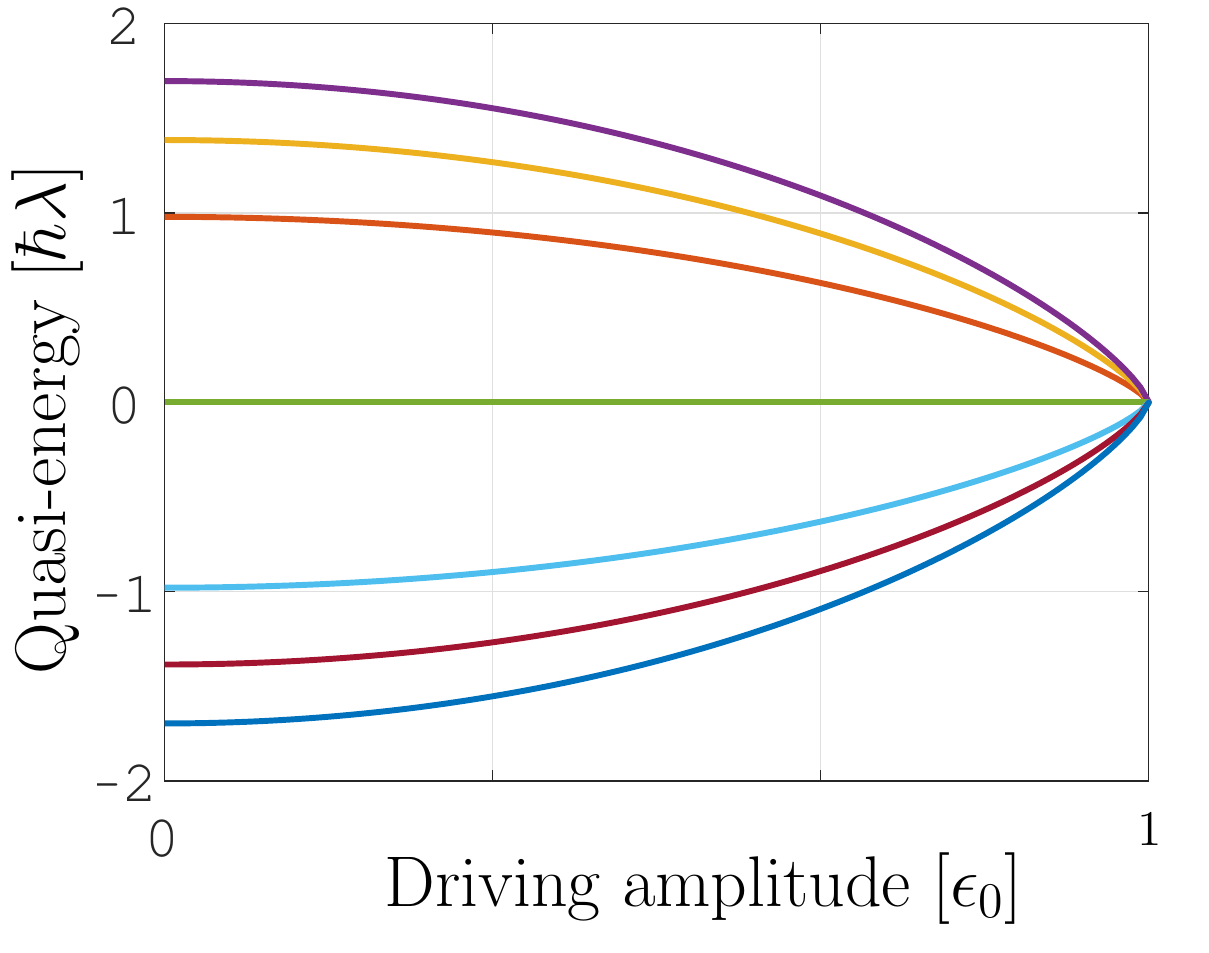}
\caption{Quasi-energy level splitting and collapse at the critical point.}\label{fig:collapse}
\end{center}
\end{figure}

The quantized motion and discrete eigenvalues found when the magnetic field dominates [equations~(\ref{eige_10}) and (\ref{eige_11a})-(\ref{eige_11b})] contrast the continuous behaviour encountered when the electric field dominates [equations~(\ref{eige_21a})-(\ref{eige_21b})]. As mentioned in the introduction, in the simplified case where $k_{2}$ and $k_{3}$ both equal zero, the $\phi^{-}_{D}$ spinor of these two complementary cases corresponds to the eigenstates of the driven Jaynes--Cummings Hamiltonian below and above critical drive. A direct comparison between equation~(\ref{back_1}) and~(\ref{corr_9}) shows that the dipole coupling constant and driving amplitude relate to the electromagnetic fields through
\begin{eqnarray}\label{qpt_1}
\lambda &= \sqrt{2}e B l_{\mathcal{B}} \, , \\
\epsilon &= e E l_{\mathcal{B}}/\sqrt{2} \, ,
\end{eqnarray}
such that the $F_{\mu \nu} F^{\mu \nu} =0$ condition refers to the transition point given in equation~(\ref{back_2}). The inclusion of $l_{\mathcal{B}}$ in these definitions already hints at the importance of the magnetic field to the existence of discrete states. It is instructive to use the above relations to obtain the discrete quasi-energies of the driven JC model from equation~(\ref{eige_10}):
\begin{equation}
\upvarepsilon_{n, \pm}^{0} = \pm \hbar \lambda \left(1 - \frac{4 \epsilon^{2}}{\lambda^{2}}  \right)^{3/4} \sqrt{n +1} \, .
\end{equation}
As mentioned above, the same result has been obtained through an algebraic method by Alsing, Guo and Carmichael (see equation~(50) of Ref.~\cite{Alsing_1992_2}) where the effect of the external driving field is interpreted as a dynamic Stark shift of the JC quasienergies. This can be extended to include counter-rotating interactions through equation~(\ref{back_4}), leading to
\begin{equation}
\upvarepsilon_{n, \pm}^{\eta} = \pm \hbar \lambda \left(1 - \frac{4 \epsilon^{2}}{(1+\eta)^{2}\lambda^{2}}  \right)^{3/4} \sqrt{n +1} \, ,
\end{equation}
and a displaced transition point. In Figure~\ref{fig:collapse}, the discrete quasienergies of the driven JC model are displayed as a function of the driving amplitude. The solutions were obtained by computing the eigenvalues of $\mathcal{H}_{0}$ in a truncated basis and show the characteristic $\sqrt{n}$-dependence and ensuing collapse when the critical point is reached.

\begin{figure}[h]
\begin{center}
\includegraphics[width=.5\linewidth]{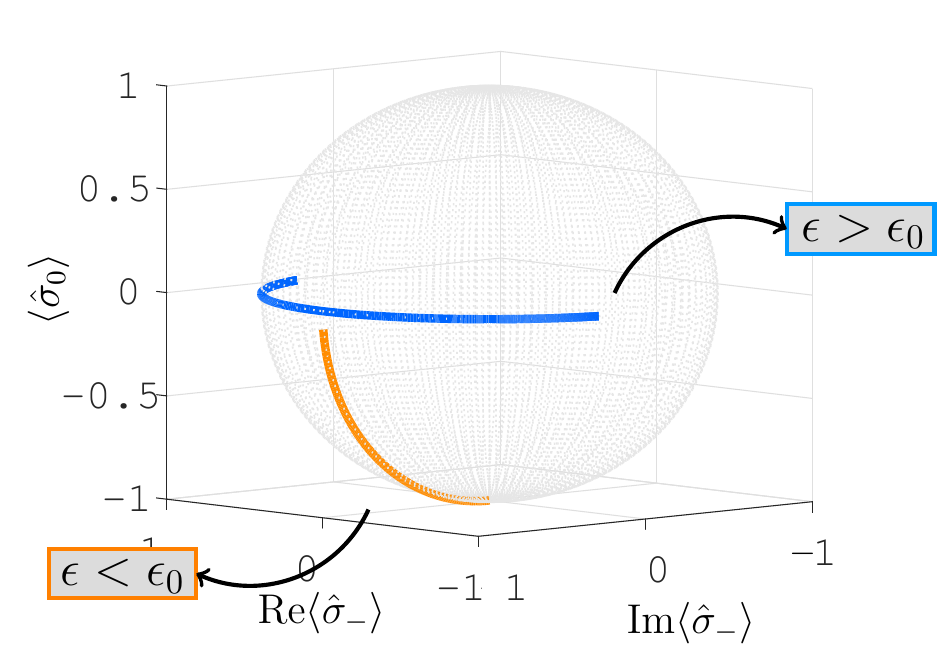}
\caption{Bloch sphere representation of the polarization $\langle \hat{\mathbf{\sigma}} \rangle$  below (orange) and above (blue) the critical point $\epsilon_{0} = \lambda /2$. Two branches arise above the critical point since $\phi^{+}_{D}$ is also a solution of the JC Hamiltonian for $\varepsilon = 0$ under $E \rightarrow - E$ [see equations~(\ref{corr_6a}) and (\ref{corr_6b})].}\label{fig:dipole}
\end{center}
\end{figure}

Besides the collapse of the level structure of the driven JC Hamiltonian, the behaviour of the zero quasienergy state captures the nature of the transition at hand. Below the critical drive amplitude, the corresponding eigenstate in the position representation is
\begin{equation}\label{qpt_2}
\phi_{D}^{-}  = \exp\left[- \frac{\sqrt{1 - \beta_{\mathcal{B}}^{2}} }{2 l_{\mathcal{B}}^{2}} x_{1}^{2} \right] \left( \begin{array}{c}
\Xi_{\mathcal{B}}^{-}   \\
-i \Xi_{\mathcal{B}}^{+} \\
\end{array} \right) \, .
\end{equation}
This separable product of field-mode and two-level system states is normalizable. The coherent drive determines the polarization of the two-level system through
\begin{equation}\label{qpt_3}
\langle \hat{\sigma}_{-} \rangle = i \beta_{\mathcal{B}} = i 2\epsilon/\lambda \, .
\end{equation}
On the other hand, above critical drive, the eigenstate is
\begin{equation}\label{qpt_4}
\phi_{D}^{-} = \exp\left[- \frac{i \sqrt{1 - \beta_{\mathcal{E}}^{2}} }{2 l_{\mathcal{E}}^{2}} x_{1}^{2} \right] \left( \begin{array}{c}
\Xi_{\mathcal{E}}^{+} + i \Xi_{\mathcal{E}}^{-}  \\
\Xi_{\mathcal{E}}^{+} - i \Xi_{\mathcal{E}}^{-}  \\
\end{array} \right) \, ,
\end{equation}
a separable product which can be delta-normalized~\cite{Lo_1990}. Disregarding the normalization factor associated with the field mode, the polarization of the two-level system can be shown to settle in the equator of the Bloch sphere with its phase depending on the drive amplitude,
\begin{equation}
\langle \hat{\sigma}_{-} \rangle = \sqrt{1- \beta_{\mathcal{E}}^{2} } + i \beta_{\mathcal{E}} = \sqrt{1- \frac{\lambda^{2}}{4\epsilon^{2}} } + i \frac{\lambda}{2 \epsilon} \, ,
\end{equation}
as displayed in Figure~\ref{fig:dipole}. We have considered the zero quasi-energy states to underline the subtle differences between states above and below the critical point. The value of the polarization is, however, only dependent of the phase of the system and remains the same for each quasi-energy state. The same polarization values are found under the semiclassical approximation when dissipation and $N$ two-level systems are considered, but with finite photon number expectations~\cite{Alsing_1992_1}. In fact, the divergences encountered above the critical point help to exemplify the role of dissipation in optical systems. Without dissipation, once the system is driven above the critical point, the field radiated by the two-level system is no longer strong enough to interfere destructively with the coherent drive [see equation~(\ref{qpt_3})]. This leads to stationary states displaying an infinite photon number. Dissipation through the cavity walls would overcome the effect of the driving field, thus avoiding the divergences in photon number. While adding dissipation to the model goes beyond the scope of the current report, detailed studies of its effects in the driven Jaynes-Cummings model are presented in Refs.~\cite{Carmichael_2015} and \cite{Alsing_1992_1}. The authors consider dissipation through two decay rates: one, $\kappa$, is related to damping of the field mode, and the other, $\gamma/2$, to spontaneous emission of the two-level system. It is found that even in the presence of these rates, the critical point identified in the Dirac-particle analogue [equation~(\ref{back_2})] remains as the organizing center of the transition. For non-vanishing $\kappa$ the system reaches a steady-state in the long-time limit with zero photon number expectation below critical drive and a finite value above it. The interplay between $\kappa, \lambda$ and $\epsilon$ determine the maximum photon value, allowing for the definition of a thermodynamical limit in this system where threshold behaviour and its relation to a phase transition can be studied. For the generalized model ($\eta\neq0$), damping of the cavity mode is considered in Ref.~\cite{Gutierrez_2017b}.

\subsection{Squeezing and Lorentz transformations}

A word is in order regarding the apparent relationship between Lorentz transformations and squeezing, as found in the solutions below critical drive. The eigenstates found in the privileged frame, equation~(\ref{eige_8b}), correspond to the dressed states of the Jaynes-Cummings model. When the Lorentz transformation was applied to reach the laboratory frame, the position and accompanying momentum operator appeared rescaled. This would translate to a squeezing transformation of these states. In reality, the natural length in the privileged frame is a relativistic invariant,
\begin{equation}
l = \sqrt{ \frac{\hbar c }{ e ( F_{\mu \nu}F^{\mu \nu} )^{1/2}}} \, ,
\end{equation}
\noindent
which rules the behaviour of the system, and in this sense the transformation does not generate squeezing. The value of this quantity also helps to explain how the behaviour of the particle reacts towards the flipping of the potential introduced between magnetic and electric field-dominated cases. A real (complex) value determines discrete (continuous) solutions.

The creation and annihilation operators given in equation~(\ref{corr_7}), however, are defined in terms of $l_{\mathcal{B}}$, not the relativistic invariant. They are affected by the Lorentz transformation. In the optical analogue these operators act upon the photon number of the field mode and the squeezing refers to photons injected into the mode through the coherent drive~\cite{Alsing_1992_2}. In the mechanical analogue they act upon the Landau states. The expected particle number in these states corresponds to the number of magnetic flux quanta $(\Phi = \hbar c / \vert e \vert)$ that can be trapped in a circle of radius $l_{\mathcal{B}}$, a radius that changes between reference frames. It is important to stress out that these operators govern the quantized motion of the system, and are not responsible of creating and annihilating particles described by the Dirac equation. When working in a flat spacetime---as in the present report---there is a unique definition of vacuum regardless of the reference frame~\cite{Ford_1998}; therefore, Lorentz transformations do not create particles and can not lead to squeezed states. This would not be the case in accelerated or curved spaces~\cite{Davies_1975,Unruh_1976,Candelas_1978,Jauregui_1991}.

\section{Conclusion}\label{secc:conclusion}

We have obtained the eigenstates and eigenvalues of the on-resonance driven Jaynes-Cummings model, and, through equations~(\ref{back_6})-(\ref{back_7_3}), the driven Jaynes-Cummings-Rabi model. The solutions display two different phases of the system depending on the ratio between dipole coupling strength and driving field amplitude. Below a critical drive amplitude, the system exhibits a discrete level structure accompanied by a finite photon number expectation. This structure collapses at the critical drive, and becomes continuous with a divergent photon number expectation above it.

The exact solutions were obtained by mapping the driven Jaynes-Cummings model onto the evolution of a Dirac particle in an electromagnetic field. The relativistic covariance of the Dirac equation was then exploited and the solutions mapped back to the quantum optical system. Under this correspondence, dipole coupling and drive amplitude are related to magnetic and electric fields.
When the former dominates, the interaction between spin and magnetic field creates a harmonic potential and discrete states naturally arise. When the latter dominates, the effective potential is unbounded from below and continuous solutions are encountered.

Aside from providing a theoretical method for finding solutions to the driven Jaynes-Cummings Rabi model, the introduced correspondence opens up a further  connection between quantum phase transitions in quantum optical and condensed matter systems.

\ack

This research was supported by the Marsden fund of the Royal Society of New Zealand. R.G.J was supported on a scholarship funded by the New Zealand Tertiary Education Committee through the Dodd-Walls Centre for Photonic and Quantum Technologies. R.G.J. thanks R.~J\'{a}uregui for insightful discussions.

\end{document}